\documentclass[12pt,reqno]{article}
\usepackage{amsfonts}
\usepackage{amsmath}
\usepackage{amsbsy}

\usepackage{amssymb,latexsym}
\numberwithin{equation}{section}

\begin{document}
 \allowdisplaybreaks[4]
\title{Integrated Field Equations of Heterotic Supergravities}
\author{Nejat T. Y$\i$lmaz\\
          Department of Mathematics
and Computer Science,\\
\c{C}ankaya University,\\
\"{O}\u{g}retmenler Cad. No:14,\quad  06530,\\
 Balgat, Ankara, Turkey.\\
          \texttt{ntyilmaz@cankaya.edu.tr}}
\maketitle
\begin{abstract}The first-order bosonic field equations of the
$D$-dimensional effective low energy theory which describes the
massless background coupling of the $D$-dimensional fully Higgsed
heterotic string are derived.
\end{abstract}

\section{Introduction}
The ten-dimensional $\mathcal{N=}1$ supergravity
\cite{d=10,tani15} which is coupled to $16$ gauge multiplets with
the gauge group either $O(32)$ or $E_{8}\times E_{8}$ is the low
energy effective limit theory which describes the massless
background coupling of the ten-dimensional heterotic string
\cite{kiritsis}. If one chooses Abelian gauge multiplets  then one
obtains the maximal torus sub-theory of the ten-dimensional
$O(32)$ or $E_{8}\times E_{8}$ Yang-Mills supergravity theory. In
this case the full gauge group is broken down to its maximal torus
subgroup $U(1)^{16}$, whose Lie algebra is the Cartan subalgebra
of the non-abelian gauge groups mentioned above. This mechanism is
due to the general Higgs vacuum structure of the heterotic string
which causes a spontaneous symmetry breakdown. Thus in this
respect upon compactification one obtains the $D$-dimensional
fully Higgsed massless heterotic string coming from the maximal
torus sub-theory of the ten-dimensional $O(32)$ or $E_{8}\times
E_{8}$ Yang-Mills supergravity. In either of the $O(32)$ or the
$E_{8}\times E_{8}$ heterotic string theories in order to obtain
the $D$-dimensional massless heterotic string in the fully Higgsed
compactification only the ten-dimensional $16$ Cartan gauge fields
are kept in the reduction since the non-cartan gauge fields lead
to massive fields following the compactification. In other words
only the Cartan gauge fields are kept since they are the only
fields which will remain massless for generic values of the Wilson
lines.

In \cite{heterotic} one can refer to the torodial compactification
of the bosonic sector of the ten-dimensional $\mathcal{N=}1$
supergravity which is coupled to $16$ Abelian gauge multiplets. As
we have discussed above such a reduction gives us the
$D$-dimensional bosonic low energy theory of the massless sector
of the $D$-dimensional fully Higgsed heterotic string. In a
formalism which treats the scalar manifolds as generic
$G/K$-cosets  and which uses the solvable Lie algebra
parametrization \cite{fre,nej1,nej2} the field equations of these
bosonic theories are studied in \cite{het}.

Starting from the bosonic field equations of the $D$-dimensional
effective massless fully Higgsed heterotic string which is the
$D$-dimensional heterotic supergravity in this note we derive the
first-order field equations by locally integrating the
second-order field equations obtained in \cite{het}. By
integration we mean cancelling an exterior derivative on both
sides of the equations. Therefore we obtain a first-order
formulation of the theory. We will effectively make use of the
results derived in \cite{consist2} which states that there exists
a one-sided on-shell decoupling between the coset scalars and the
gauge fields of the heterotic supergravities. For this reason to
obtain the first-order field equations of the coset scalars we
will adopt the general formulation of \cite{nej2} which works out
the first-order field equations of the pure symmetric space sigma
model. We will also give a brief discussion how one can make use
of the first-order field equations to perform the on-shell bosonic
coset construction of the $D$-dimensional heterotic supergravity.
\section{The First-Order Field Equations}\label{section23}
The bosonic field content which constitutes the low energy
effective Lagrangian that describes the bosonic sector of the
 $D$-dimensional massless background coupling of the
fully Higgsed heterotic string can be given as \cite{het}
\begin{equation}\label{eq1}
\{C_{(1)}^{I},A_{(2)},\phi,\phi^{i},\chi^{\alpha}\}.
\end{equation}
The fields $C_{(1)}^{I}$ are $(20-2D+16)$ one-forms, and $A_{(2)}$
is a two-form, the rest of the fields are scalars. The scalar
field $\phi$ is decoupled from the rest of the scalars which are
the coset ones. The coset scalars $\{\phi^{i},\chi^{\alpha}\}$
parametrize the coset manifold
\begin{equation}\label{eq2}
O^{\prime}(10-D+16,10-D)/O(10-D+16)\times O(10-D),
\end{equation}
whose elements which are ($20-2D+16$)-dimensional real matrices in
the fundamental representation satisfy
\begin{equation}\label{eq3}
\nu^{T}\Omega \nu=\Omega.
\end{equation}
Here the $(20-2D+16)\times(20-2D+16)$ matrix $\Omega$ is
\begin{equation}\label{eq4}
\Omega=\left(\begin{array}{ccc}
  0 & 0 & -{\mathbf{1}}_{(10-D)} \\
  0 & {\mathbf{1}}_{(16)} & 0 \\
  -{\mathbf{1}}_{(10-D)} & 0 & 0 \\
\end{array}\right),
\end{equation}
where ${\mathbf{1}}_{(n)}$ is the $n\times n$ unit matrix.
Following the notation of \cite{het} we use a prime in \eqref{eq2}
which stands for the particular representation of
$O(10-D+16,10-D)$ defined through \eqref{eq3} that is generated by
the indefinite signature metric \eqref{eq4}\footnote{On the other
hand the unprimed notation $O(10-D+16,10-D)$ is used in \cite{het}
for the usual representation of the generalized orthogonal group
generated by the diagonalized indefinite signature metric
$\eta=\text{diag}(-,-,...,-,+,+,...,+)$.}. We should also state
that we separate $N=16$ in our expressions to emphasize the number
of Abelian matter multiplets coupling to the ten-dimensional
$\mathcal{N=}1$ type I supergravity which forms when coupled to
these $16$ $U(1)$ vector multiplets the low energy effective limit
of the ten-dimensional fully Higgsed heterotic string. The coset
parametrization can be constructed in the solvable Lie algebra
gauge \cite{fre,nej1,nej2,het} as
\begin{equation}\label{eq5}
\nu=e^{\frac{1}{2}\phi ^{i}H_{i}}e^{\chi ^{\alpha}E_{\alpha}},
\end{equation}
where $i=1,\cdots,r$ and $\alpha=1,\cdots,n$. Here $H_{i}$ are the
Cartan generators and $E_{\alpha}$ are the positive root
generators of the solvable Lie algebra which takes part in the
Iwasawa decomposition of $o^{\prime}(10-D+16,10-D)$ \cite{hel}.
One can find a more detailed study of the solvable Lie algebra
parametrization in \cite{nej1,nej2,het}. The field equations of
the bosonic fields \eqref{eq1} are already derived in \cite{het}.
They read
\begin{equation}\label{eq6}
\begin{aligned}
(-1)^{D}d(\ast
d\phi)&=\frac{1}{2}\sqrt{8/(D-2)}\:e^{-\sqrt{\frac{8}{(D-2)}}\phi}
\ast F_{(3)}\wedge
F_{(3)}\\
&\quad
+\frac{1}{2}\sqrt{2/(D-2)}\:e^{-\sqrt{\frac{2}{(D-2)}}\phi}\mathcal{M}_{IJ}
\ast H_{(2)}^{I}\wedge H_{(2)}^{J},\\
d(e^{-\sqrt{\frac{8}{(D-2)}}\phi}\ast F_{(3)})&=0,\\
d(e^{-\sqrt{\frac{2}{(D-2)}}\phi}\mathcal{M}^{I}_{\:\:\:J}\ast
H_{(2)}^{J})&=(-1)^{D}e^{-\sqrt{\frac{8}{(D-2)}}\phi}\Omega^{I}_{\:\:\:J} H_{(2)}^{J}\wedge\ast F_{(3)},\\
d(e^{\gamma _{i}\phi ^{i}}\ast U^{\gamma })&=\sum\limits_{\alpha
-\beta=-\gamma }N_{\alpha ,-\beta }U^{\alpha }\wedge e^{\beta
_{i}\phi ^{i}}\ast U^{\beta },\\
d(\ast d\phi ^{i})&=\frac{1}{2}\sum\limits_{\beta\in\Delta_{nc}^{+}}^{}\beta _{i}%
e^{\frac{1}{2}\beta _{j}\phi ^{j}}U^{\beta }\wedge e^{%
\frac{1}{2}\beta _{j}\phi ^{j}}\ast U^{\beta
}\\
&\quad -\frac{1}{2}(-1)^{D}e^{-\sqrt{\frac{2}{(D-2)}}\phi}\ast
H_{(2)}\wedge\nu^{T}H_{i}\nu H_{(2)},
\end{aligned}
\end{equation}
where $\alpha,\beta,\gamma$ whose corresponding generators enter
in the solvable lie algebra parametrization of \eqref{eq5} are the
elements of $\Delta_{nc}^{+}$ which is the set of non-compact
positive roots of $o^{\prime}(10-D+16,10-D)$
\cite{nej2,het}\footnote{We adopt the notation of \cite{nej2} and
randomly enumerate the roots in $\Delta_{nc}^{+}$ from $1$ to
$n$.}. The field strengths of the fields
$\{C_{(1)}^{I},A_{(2)},\chi^{\beta}\}$ are respectively defined as
\begin{equation}\label{eq7}
\begin{aligned}
H_{(2)}^{I}&=dC_{(1)}^{I},\\
F_{(3)}&=dA_{(2)}+\frac{1}{2}\:\Omega_{IJ}\: C_{(1)}^{I}\wedge
dC_{(1)}^{J},\\
U^{\alpha}&=\mathbf{\Omega}^{\alpha}_{\:\:\:\beta}d\chi^{\beta}.
\end{aligned}
\end{equation}
The matrix $\mathcal{M}$ is
\begin{equation}\label{eq7.5}
\mathcal{M}=\nu ^{T}\nu.
\end{equation}
In the above relations $\beta_{i},\gamma_{i}$ are the root vector
components and $N_{\alpha,\beta}$ are the structure constants of
the corresponding positive root generators of the solvable Lie
algebra generated by $\{H_{i},E_{\alpha}\}$ \cite{nej1,nej2,het}.
More specifically
\begin{equation}\label{yeni0.5}
[H_{j},E_{\gamma}]=\gamma_{j}E_{\gamma},
\end{equation}
and
\begin{equation}\label{yeni0.6}
[E_ {\alpha},E_{\beta}]=N_{\alpha,\beta}E_{\alpha+\beta}.
\end{equation}
From \cite{nej1,nej2} the definition of the $n\times n$ matrix
$\mathbf{\Omega}(\chi^{\beta})$ reads\footnote{The reader should
be aware that the plain $\Omega$ with indices $I,J,K,...$ and the
bold one $\mathbf{\Omega}$ with the indices
$\alpha,\beta,\gamma,...$ are two different objects. We prefer
using them together for the sake of conformity with the
references.}
\begin{equation}\label{yeni1}
 \mathbf{\Omega}=(e^{\omega}-I)\,\omega^{-1},
\end{equation}
where $\omega$ is an $n \times n$ matrix with components
\begin{equation}\label{yeni2}
\omega_{\beta}^{\gamma}=\chi^{\alpha}K_{\alpha\beta}^{\gamma}.
\end{equation}
Here $K_{\alpha\beta}^{\gamma}$ are defined through the
commutators of $\{E_{\alpha}\}$\footnote{Of course by comparing
\eqref{yeni3} with \eqref{yeni0.6} one may relate
$N_{\alpha,\beta}$ to $K_{\alpha\beta}^{\gamma}$.}
\begin{equation}\label{yeni3}
[E_{\alpha},E_{\beta}]=K_{\alpha\beta}^{\gamma}E_{\gamma}.
\end{equation}
We should state that we freely lower and raise indices by using
various dimensional Euclidean metrics when necessary for
convenience of notation. In \cite{het} it is shown that the second
term on the right hand side of the last equation in \eqref{eq6}
which is compactly written in matrix form can be given as
\begin{equation}\label{eq8}
-\frac{1}{2}(-1)^{D}e^{-\sqrt{\frac{2}{(D-2)}}\phi}\ast
H_{(2)}\wedge\nu^{T}H_{i}\nu
H_{(2)}=(-1)^{D}\frac{\partial\mathcal{L}_{m}}{\partial\phi^{i}}.
\end{equation}
Here $H_{i}$ are the ($20-2D+16$)-dimensional matrix
representatives of the Cartan generators and $H_{(2)}$ is the
vector of the field strengths defined in \eqref{eq7}. Also
$\mathcal{L}_{m}$ is the matter-scalar coupling Lagrangian
\cite{het}. However on the other hand in \cite{consist2} it is
proven that the expression in \eqref{eq8} vanishes on-shell for
the elements of the solution space indicating that the coset
scalar field equations coincide with the pure sigma model ones.
Therefore we can legitimately drop the second term on the right
hand side of the last equation in \eqref{eq6}.

As discussed in the Introduction our aim in this note is to
integrate the field equations in \eqref{eq6} locally. In this
respect we will use the fact that locally a closed differential
form is an exact one. For this reason we first introduce the dual
fields
\begin{equation}\label{eq9}
\{\widetilde{C}^{I},\widetilde{B},\widetilde{\phi}\},
\end{equation}
where $\widetilde{C}^{I}$ are $(D-3)$-forms, $\widetilde{B}$ is a
$(D-4)$-form and $\widetilde{\phi}$ is a $(D-2)$-form. It is a
straightforward operation that if one applies the exterior
derivative on both sides of
\begin{equation}\label{eq10}
e^{-\sqrt{\frac{8}{(D-2)}}\phi}\ast F_{(3)}=d\widetilde{B},
\end{equation}
one obtains the second equation in \eqref{eq6}. Thus \eqref{eq10}
is our first first-order field equation. Next let us consider
\begin{equation}\label{eq11}
e^{-\sqrt{\frac{2}{(D-2)}}\phi}\mathcal{M}^{I}_{\:\:\:J}\ast
H_{(2)}^{J}=(-1)^{D}(d\widetilde{C}^{I}+\Omega^{I}_{\:\:\:K}C^{K}_{(1)}\wedge
d\widetilde{B}).
\end{equation}
If we take the exterior derivative of both sides we get
\begin{equation}\label{eq12}
d(e^{-\sqrt{\frac{2}{(D-2)}}\phi}\mathcal{M}^{I}_{\:\:\:J}\ast
H_{(2)}^{J})=(-1)^{D}\Omega^{I}_{\:\:\:K}dC^{K}_{(1)}\wedge
d\widetilde{B}.
\end{equation}
By using \eqref{eq10} in it \eqref{eq12} gives the third equation
in \eqref{eq6}. Finally let us take the first-order equation
\begin{equation}\label{eq13}
\ast
d\phi=d\widetilde{\phi}-\frac{1}{2}\sqrt{8/(D-2)}A_{(2)}\wedge
d\widetilde{B}+\frac{1}{2}\sqrt{2/(D-2)}\delta_{IJ}C^{I}_{(1)}\wedge
d\widetilde{C}^{J}.
\end{equation}
If we apply the exterior derivative on both sides of \eqref{eq13}
we get
\begin{equation}\label{eq14}
d(\ast d\phi)= -\frac{1}{2}\sqrt{8/(D-2)}dA_{(2)}\wedge
d\widetilde{B}+\frac{1}{2}\sqrt{2/(D-2)}\delta_{IJ}dC^{I}_{(1)}\wedge
d\widetilde{C}^{J}.
\end{equation}
By using \eqref{eq7} and \eqref{eq10} the above equation can be
written as
\begin{subequations}\label{eq15}
\begin{align}
d(\ast
d\phi)&=\frac{1}{2}\sqrt{8/(D-2)}\:e^{-\sqrt{\frac{8}{(D-2)}}\phi}
(-1)^{D}\ast F_{(3)}\wedge
F_{(3)}\notag\\
 &\quad +(-1)^{D}\frac{1}{2}\sqrt{2/(D-2)}\:e^{-\sqrt{\frac{2}{(D-2)}}\phi}\mathcal{M}_{IK}
\ast H_{(2)}^{K}\wedge H_{(2)}^{I}\notag\\
&\quad
+\frac{1}{2}\sqrt{2/(D-2)}\:e^{-\sqrt{\frac{8}{(D-2)}}\phi}\Omega_{IJ}C^{I}_{(1)}\wedge
H_{(2)}^{J}\wedge
\ast F_{(3)}\notag\\
&\quad
-\frac{1}{2}\sqrt{2/(D-2)}\:e^{-\sqrt{\frac{8}{(D-2)}}\phi}\Omega_{KI}C^{K}_{(1)}\wedge
H_{(2)}^{I}\wedge \ast üF_{(3)}.\tag{\ref{eq15}}
\end{align}
\end{subequations}
Since $\Omega$ is a symmetric matrix the last two terms cancel and
as $\mathcal{M}$ is also a symmetric matrix this equation gives us
the first equation in \eqref{eq6}. The first-order formulation of
the coset scalar field equations in \eqref{eq6} is a
straightforward task. Following our discussion above when we drop
the second term on the right hand side of the last equation in
\eqref{eq6} we obtain the pure sigma model field equations which
are the same with the ones derived for a generic coset manifold in
\cite{nej2}. As we have remarked before this fact is a consequence
of the on-shell conditions satisfied by the general solutions of
the theory which are rigorously derived in \cite{consist2}. The
first-order field equations of the general non-split \cite{hel}
scalar coset are already derived in \cite{nej2}. In general the
coset manifolds in \eqref{eq2} are also in non-split form.
Therefore we can adopt the results of \cite{nej2} for the scalar
sectors of the heterotic supergravities. For the sake of
completeness we will repeat the first-order scalar field equations
of \cite{nej2} here. From \cite{nej2} we have
\begin{equation}\label{eq16}
\ast \overset{\rightharpoonup }{\mathbf{\Psi }}=(-1)^{D}e^{\mathbf{\Gamma }%
}e^{\mathbf{\Lambda }}\overset{\rightharpoonup }{\mathbf{A}}.
\end{equation}
Here we define the $(r+n)$-dimensional column vectors
$\overset{\rightharpoonup }{\mathbf{\Psi }}$ and
$\overset{\rightharpoonup }{\mathbf{A}}$ whose components can be
given as
\begin{subequations}\label{eq17}
\begin{gather}
\mathbf{\Psi}^{i}=\frac{1}{2}d\phi^{i},\quad \text{for}\quad
i=1,...,r,\quad
\mathbf{\Psi}^{\alpha+r}=e^{\frac{1}{2}\alpha_{i}\phi^{i}}\mathbf{\Omega
}_{\:\:\:\gamma}^{\alpha}d\chi^{\gamma},\quad\text{for}\quad\alpha=1,...,n,\notag\\
\notag\\
\mathbf{A}^{i}=\frac{1}{2}d\widetilde{\phi}^{i},\quad\text{for}\quad
i=1,...,r,\quad\text{and}\quad
\mathbf{A}^{\alpha+r}=d\widetilde{\chi}^{\alpha},\quad\text{for}\quad\alpha=1,...,n,
 \tag{\ref{eq17}}
\end{gather}
\end{subequations}
where we have introduced the dual $(D-2)$-forms
$\widetilde{\phi}^{i}$ and $\widetilde{\chi}^{\alpha}$. In
\eqref{eq16} $\mathbf{\Gamma }(\phi^{i})$ and $\mathbf{\Lambda
}(\chi^{\beta})$ are $(n+r)\times(n+r)$ matrix functions. Their
components read
\begin{equation}\label{yeni7}
\mathbf{\Gamma }%
_{n}^{k}=\frac{1}{2}\phi ^{i}\,\widetilde{g}_{in}^{k}\quad,\quad
\mathbf{\Lambda }_{n}^{k}=\chi ^{\alpha}\widetilde{f}_{\alpha
n}^{k}.
\end{equation}
The real constant coefficients $\{\widetilde{g}_{in}^{k}\}$ and
$\{\widetilde{f}_{\alpha n}^{k}\}$ are already listed in
\cite{nej1}\footnote{In \cite{nej1} the indices $i,j,...$ are
taken to run from $1$ to $l$ however in the present manuscript we
have preferred using $r$ instead of $l$. Thus the reader may read
the coefficients $\{\widetilde{g}_{in}^{k}\}$ and
$\{\widetilde{f}_{\alpha n}^{k}\}$ from equations (3.8) and (3.9)
of \cite{nej1} by replacing $l$ with $r$.}. They are
\begin{subequations}\label{yeni8}
\begin{gather}
\widetilde{f}_{\alpha m}^{n}=0,\quad\quad m\leq r\quad,\quad
\widetilde{f}_{\alpha ,\alpha +r}^{i}=\frac{1}{4}\alpha _{i},\quad\quad%
i\leq r,\notag\\
\notag\\
\widetilde{f}_{\alpha ,\alpha +r}^{i}=0,\quad\quad i>r\quad,\quad
\widetilde{f}_{\alpha ,\beta +r}^{i}=0,\quad\quad i\leq r,%
\quad\alpha \neq \beta ,\notag\\
\notag\\
\widetilde{f}_{\alpha ,\beta +r}^{\gamma +r}=N_{\alpha ,-\beta
},\quad\quad \alpha -\beta =-\gamma,\quad
\alpha \neq \beta,\notag\\
\notag\\
\widetilde{f}_{\alpha ,\beta +r}^{\gamma +r}=0,\quad\quad \alpha
-\beta \neq -\gamma,\quad \alpha \neq \beta,\tag{\ref{yeni8}}
\end{gather}
\end{subequations}
and
\begin{subequations}\label{yeni9}
\begin{gather}
\widetilde{g}_{im}^{n}=0,\quad\quad m\leq r\quad,\quad\widetilde{%
g}_{im}^{n}=0,\quad\quad m>r,\quad m\neq n,\notag\\
\notag\\
\widetilde{g}_{i\alpha }^{\alpha }=-\alpha _{i},\quad\quad\alpha
>r.\tag{\ref{yeni9}}
\end{gather}
\end{subequations}
Since as discussed in detail in \cite{nej1,nej2} beside being
enumerated $\alpha,\beta,\gamma,...$ correspond to the set of
non-compact positive roots of $o^{\prime}(10-D+16,10-D)$ the
conditions on them in \eqref{yeni8} and \eqref{yeni9} must be
understood in the root sense. We should also state that likewise
in \cite{nej2} we assume the signature of the spacetime as $s=1$.
It is proven in \cite{nej2} that as a consequence of the
dualisation of the general symmetric space sigma model the
first-order field equations in \eqref{eq16} correspond to the
local integration of the last two equations of \eqref{eq6} when
the term which comes from the scalar-matter coupling Lagrangian is
dropped as discussed before. Therefore we have derived the entire
set of first-order field equations which are obtained by locally
cancelling an exterior derivative on both sides of the equations
in \eqref{eq6}. Namely the equations \eqref{eq10}, \eqref{eq11},
\eqref{eq13}, and \eqref{eq16} represent the first-order
formulation of the $D$-dimensional low energy massless background
coupling of the fully Higgsed heterotic string which is the
$D$-dimensional heterotic supergravity.

Before concluding we will present a discussion of an important
application of the first-order field equations of the heterotic
supergravities. In \cite{julia2} the locally integrated
first-order bosonic field equations of the maximal and IIB
supergravities are used to derive the superalgebras that lead to
the complete coset constructions of the bosonic sectors of these
theories. Similarly the methodology of \cite{julia2} can be
extended to the heterotic supergravities. We will not present the
complete coset construction of the heterotic supergravities here
and leave it to a future work however we will discuss the outline
of deriving the superalgebra of the on-shell coset construction of
the heterotic supergravities. The first task in constructing the
coset formalism is to assign an algebra generator to each original
and dual field in the first-order equations \eqref{eq10},
\eqref{eq11}, \eqref{eq13}, \eqref{eq16} and then to propose a
coset map. In our case this map becomes
\begin{eqnarray}\label{yeni23}
\nu &=exp(\frac{1}{2}\phi^{j}H_{j})exp(\chi^{m}E_{m})exp(\phi
K)exp(C_{(1)}^{I}V_{I})exp(\frac{1}{2}A_{(2)}Y)\nonumber\\
\nonumber\\&\quad\times
exp(\frac{1}{2}\widetilde{B}\widetilde{Y})exp(\widetilde{C}^{I}\widetilde{V}_{I})exp(\widetilde{\phi}
\widetilde{K})exp(\widetilde{\chi}^{m}\widetilde{E}_{m})exp(\frac{1}{2}\widetilde{\phi}^{j}\widetilde{H}_{j})\label{32}.
\end{eqnarray}
The associated Cartan-form may be defined as
\begin{equation}\label{yeni33}
\mathcal{G}=d\nu\nu^{-1}.
\end{equation}
From \cite{julia2} we know that in the doubled formalism coset
construction the Cartan-form satisfies a twisted self-duality
equation
\begin{equation}\label{yeni43}
\ast\mathcal{G}=\mathcal{SG},
\end{equation}
with $\mathcal{S}$ being a pseudo-involution of the coset algebra
of the generators introduced in \eqref{yeni23}. The key ingredient
of the coset construction is  the requirement that \eqref{yeni43}
must give us the first-order field equations of the theory.
Therefore the method of revealing the coset algebra structure is
to calculate \eqref{yeni33} in terms of the desired structure
constants, then to insert it in \eqref{yeni43} and finally to
compare the result with the equations \eqref{eq10}, \eqref{eq11},
\eqref{eq13}, \eqref{eq16} to read the structure constants of the
coset algebra.
\section{Conclusion}
In this work, by locally integrating the second-order field
equations which are derived in \cite{het} and which govern the
massless sector of the $D$-dimensional fully Higgsed heterotic
string namely the $D$-dimensional heterotic supergravity we have
obtained the first-order field equations of the theory which
contain only a single exterior derivative acting on the
potentials. In these first-order field equations we have
introduced dual fields which may be considered as integration
constants. The dual fields are nothing but the Lagrange
multipliers associated with the Bianchi identities of the field
strengths when one treats these field strengths as fundamental
fields instead of their potentials \cite{pope}. The fact which is
derived in \cite{consist2} that as an on-shell condition the coset
scalar field equations can completely be decoupled from the gauge
fields provides us the usage of the first-order symmetric space
sigma model field equations of \cite{nej2} in our formulation.

In GR the Palatini application of the Ostrogradski method
\cite{ostro} of reducing the derivative order of second-order
Lagrangians by including auxiliary fields is a vast research area
in recent years especially for the f(R) theories of gravity. The
Ostrogradski method have also been effectively used to obtain the
first-order formulations of supergravity theories. First-order
formulations of supergravities are studied to understand the
superpotentials \cite{pot1,pot2,pot3} as well as the supersymmetry
transformation laws \cite{juliasilva}. The reader may find
examples of the first-order formalism of supergravity theories in
various dimensions in
\cite{juliasilva,first1,first2,first3,first4,first5,first6}. In
the general first-order formalism method of these works the field
strengths of the basic fields are also considered as independent
fields and a first-order Lagrangian is constructed which gives
first-order field equations in terms of the basic fields and their
field strengths. When the field equations of the field strengths
are substituted back in the Lagrangian one recovers the
second-order formalism. In comparison with this scheme our
first-order field equations of the $D$-dimensional heterotic
supergravity do contain the basic fields except the graviton but
on the contrary they do not include the field strengths. Instead
we have introduced dual fields which may be considered as
arbitrary integration constants that algebraically came into the
scene as a result of abolishing an exterior derivative on both
sides of the second-order field equations. Thus our approach is
purely algebraic rather than being formal. We have simply reduced
the degree of the field equations without increasing the number of
fields to be solved and in this process arbitrary integration
constants have arouse. On the other hand we have not constructed
the corresponding Lagrangian which would lead to the first-order
equations we have obtained. However as we have discussed above
such a Lagrangian which would kinematically be different than the
one that would appear within the Ostrogradski method would rather
be obtained by Lagrange multiplier method that makes use of the
Bianchi identities of the field strengths.

The first-order formulation of the $D$-dimensional heterotic
supergravity presented in this note has two important
implications. The first-order field equations play an important
role in the coset construction of the supergravities
\cite{julia2}. Thus as we have briefly discussed in the previous
section the equations derived in this note can be considered to be
essential ingredients of a possible coset construction of the
heterotic supergravities. Secondly since the dual fields
introduced in the first-order field equations can be arbitrarily
varied one can make use of this fact to generate solutions.
Therefore in this respect beside being first-order the integrated
field equations containing parameters which can be manipulated
become powerful tools in seeking solutions of the heterotic
supergravities.

\end{document}